\documentclass[floatfix,twocolumn,amsmath,amssymb,aps]{revtex4}
\usepackage{graphicx}
\usepackage{dcolumn}
\usepackage{bm}
\usepackage{siunitx}
\usepackage{color}
\usepackage{ulem}

\begin{document}

\preprint{APS/123-QED}

\title{Energy dissipation of a sphere rolling up a granular slope: slip and deformation of granular surface}

\author{T. Fukumoto, K. Yamamoto, M. Katsura and H. Katsuragi}
\affiliation{
 Department of Earth and Space Science, Osaka University, 1-1 Machikaneyama, Toyonaka 560-0043, Japan
 }

\date{\today}

\begin{abstract}
We experimentally investigate the dynamics of a sphere rolling up a granular slope. During the rolling-up motion, the sphere experiences slipping and penetration (groove formation) on the surface of the granular layer. The former relates to the stuck motion of the rolling sphere, and the latter causes energy dissipation due to the deformation of the granular surface. To characterize these phenomena, we measured the motion of a sphere rolling up a granular slope of angle $\alpha$. The initial velocity $v_0$, initial angular velocity $\omega_0$, angle of slope $\alpha$, and density of the sphere $\rho_s$ were varied. As a result,
the penetration depth can be scaled solely by the density ratio between the sphere and granular layer. 
By considering the rotational equation of motion, we estimate the friction due to the slips. Besides, by considering energy conservation, we define and estimate the friction due to groove formation. Moreover, the translational friction is proportional to the penetration depth. Using these results, we can quantitatively predict the sphere's motion including stuck behavior.
\end{abstract}

\maketitle

\section{\label{sec:intro}INTRODUCTION}
In general, granular bed is so deformable that various phenomena such as impact cratering~\cite{katsuragi_2016} and wedge formation by plowing ~\cite{Gravish:2010} can be induced. Granular deformability enables us to draw even sand art~\cite{sandart_2017}. We focus on a certain type of deformation and related frictional phenomenon occurring on the granular surface. Specifically, on loose sand surfaces, vehicles might be stuck by the wheels spinning out. Preventing such stuck behavior is a key of efficient vehicle design. The wheel-stuck phenomena can induce severe malfunction of the planetary explorator as well. For example, the Mars rover, Spirit, was stuck on the surface of Mars. After that, Spirit could not continue the exploration of the Martian environment~\cite{Sanderson}. Efficient vehicle design is the main topic in the field of  terramechanics research. Particularly, the slip ratio of the rover's wheel that depends on the shape and driving conditions of the rover has been studied extensively~\cite{sutoh1,nagaoka1}. The principal goal of these terramechanics researches is developing high-performance rovers. Thus, the specific geometry and setup of the rover have been studied. 

However, a fundamental understanding of the relationship between a granular surface and a simple object (such as a sphere) has not been sufficient. Investigation of such a fundamental relation could also relate to the ecology of antlions that use the stuck phenomena to prey on ants~\cite{ant, Crassous:2017}.  
It would also relate to groove formation on planetary granular surfaces. For example, traces of boulder falls on the regolith layer have been found~\cite{Kumar:2016,Ikeda:2022}. To discuss the mechanics of such boulder-fall traces, the interaction between a granular surface and a macroscopic object must be revealed. Namely, while the setup we consider --- rolling sphere on a granular slope --- is quite simple, it relates to various phenomena such as vehicle design (engineering), ecology of antlions, and planetary surface processes.  

For a proper understanding of the interaction between a solid object and a granular surface, frictional property is the most important factor. Since granular frictional behaviors are quite diverse and complex, various efforts have been made to reveal the constitutive law of granular friction~(e.g., \cite{Pouliquen2006,daCruz2005,Kuwano2013}). Regarding the interaction between granular matter and objects, various friction-related phenomena have also been investigated (e.g., friction during plowing~\cite{Gravish:2010}, penetration~\cite{Schroter:2007}, and withdrawing~\cite{Furuta:2019}). Through these studies, our understanding of granular friction has been developed. However, these researches have not analyzed the combination of translational and rotational friction.

In addition, some researchers examined frictional drag force exerting on an intruder in a granular bed~\cite{F.Pacheco:2009,Theodore:2010,V.L.Diaz:2020,Shashank:2023}. These researches clarified the friction (drag force) of translational and rolling motion. However, they focused on the drag force within a bulk granular bed. Thus, the details of drag force exerting on an object (sphere) rolling on a granular free surface have not been revealed by these studies.

Some experiments rolling a sphere on a granular surface have been carried out. For example, the dynamics of a sphere rolling down on an inclined rough surface on which grains were glued has been examined~\cite{effects,energy,model}.
Measurement of the friction coefficient
has also been performed for the sphere rolling down on a free granular surface. According to \cite{Blasio}, the friction coefficient characterizing energy dissipation is mainly dependent on the sphere's density and almost independent of the motion velocity.
This research revealed that the energy dissipation is mainly caused by the sinking of the sphere object. However, a quantitative evaluation of energy dissipation due to the sinking and slipping during the motion has not been carried out. When a basketball or a medicine ball rolls on the granular surface,
the friction coefficient characterizing the rolling resistance was estimated by experiments and numerical simulations 
\cite{Stefaan}. However, this research was conducted under the no-slip situation, so the stuck phenomenon could not be captured. Recently, Texier et al.  studied the motion of a sphere rolling down on an inclined granular surface~\cite{Melo}. When a sphere was placed on a granular slope, the sphere's behavior depended on the slope angle and density ratio between the sphere and granular matter. They mainly focused on the accelerative motion of the sphere. However, the deceleration of the sphere rolling up a granular slope has not been studied. While the motivation and setup of this study are quite simple, any systematic experiment on the motion of the sphere rolling up a granular slope has not been performed.

Here, we experimentally examine the sphere rolling up a free granular slope in order to characterize the decelerative motion and to quantify energy dissipation caused by slip motion and groove deformation. To simply discuss energy conservation, rolling-up motion without external driving force is investigated. Although this setup is different from the actual wheel driving situation, the stuck phenomenon, which is characterized by sphere spinning without translational motion, can be mimicked by this simple setup. Using this setup, we characterize the passive (without driving force) stuck motion.
We distinguish the friction due to the slipping and shallow sinking (groove formation) of the sphere. And finally, the form to predict rolling-up dynamics is obtained based on energy conservation.

In the next section, the experimental setup, parameters, and other conditions are described. Then, the experimental results (dynamics of the sphere rolling up the slope) are presented in Sec.~\ref{sec:results} and analyzed in Sec.~\ref{sec:analysis}. After discussing the physical meaning of the obtained results in Sec.~\ref{sec:discussion}, the conclusion is provided in Sec.~\ref{sec:conclusion}. 

\section{EXPERIMENT} \label{sec:experiment}
\begin{figure} 
\includegraphics[width=\columnwidth]{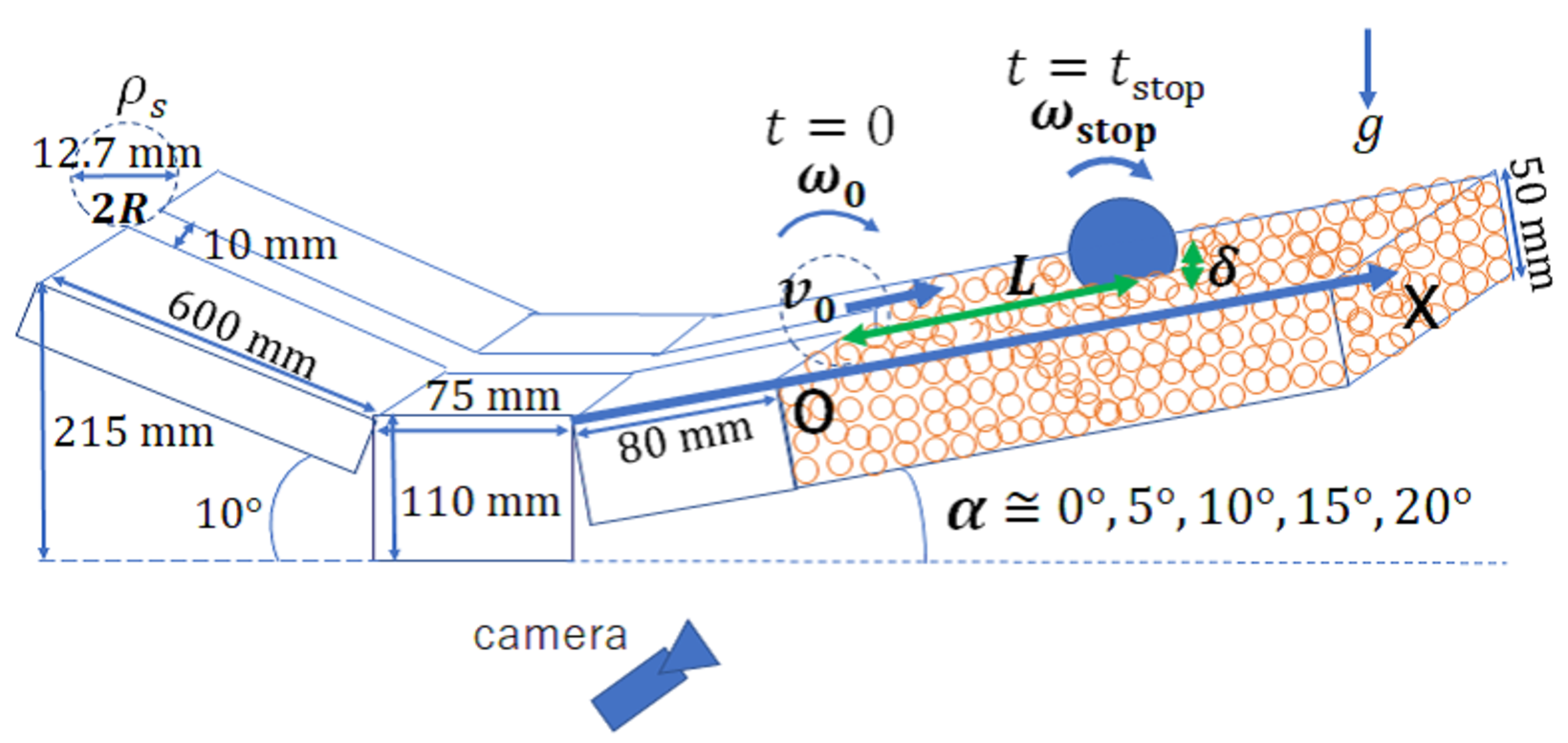} 
\caption{Experimental setup. By rolling down the slope of the rail made of aluminium, the sphere obtains initial translational velocity $v_0$ and angular velocity $\omega_0$ when it enters the granular slope at $X=0$. The sphere rolls up the granular slope from $t=0$ to $t=t_\mathrm{stop}$ (and $X=L$). The $X$ axis is defined along the slope surface. The slope angle is varied from $\alpha \simeq 0{^\circ}$ to $20^{\circ}$.}
\label{fig:exp_sys}
\end{figure}

To measure the dynamics of a sphere rolling up a granular slope, we build an experimental apparatus as schematically shown in Fig.~\ref{fig:exp_sys}. A sphere is released at a certain height ($160$, $180$, $200$, or $215$~mm from the base level) on the rail made of aluminum (rail width:~$10$~mm). The spheres used in this experiment are made of polyethylene, polyacetal, glass, alumina ceramic, and stainless steel. All these spheres have an identical radius, $R=6.35$~mm. Their densities are $\rho_{s}= 930$, $1400$, $2600$, $3900$, and $7900$~kg/m$^{3}$, respectively. After rolling down/up the rail, the sphere enters into a glass-beads layer at $X=0$ at time $t=0$. The $X$ axis is taken along the surface of the granular layer.
The inclination angle $\alpha$ is constant in the region of $X \geq -80$~mm. The sphere enters the granular layer with an initial translational velocity $v_0$($=dX/dt$ at $t=0$) and initial angular velocity $\omega_0$. 
Typical diameter of the glass beads used in this experiment is $0.8$~mm (\SI{770}{\micro m} - \SI{910}{\micro m}, AS-One, BZ-08) and the thickness of the granular layer is $50$~mm. Granular bulk density in this experiment is estimated as $\rho_{g}=~1560\pm~20$~kg/m$^{3}$ which corresponds to the packing fraction of $\varphi = 0.64 ~\pm~0.01.$ The inclination angle of the granular slope $\alpha$ is varied as $\alpha\simeq 0^{\circ}$, $5^{\circ}$, $10^{\circ}$, $15^{\circ}$, and $20^{\circ}$. By varying $\alpha$, the effect of gravity on the motion of the sphere changes. The effect of $\alpha$ on translational and rotational slip motions has not been revealed in previous studies. To predict the rolling deceleration under various $\alpha$ conditions, systematic experiments are necessary. Through the systematic experiments conducted in this study, the degree of stuck motion can be characterized.

By varying the releasing position of the sphere, $v_{0}$ (and $\omega_0$) can be controlled. Experimental runs with identical conditions are repeated five times to check the reproducibility. Before each experimental run, the granular layer is completely refreshed to ensure the homogeneously flat initial condition. In specific, we remove all glass beads from the box after each experimental run and pour them again. Then, the flat surface is maintained. The mass of glass beads in the box is 1.68 $\pm$ 0.02~kg, resulting in $\varphi = 0.64 ~\pm~0.01$. The dynamics of the sphere's motion during the rolling-up process was captured by a high-speed camera (Omron Sentech, STC-MBS163U3V) with $200$~fps and $0.21$~mm/pixel resolution (image size:~1,080$\times$1,440 pixels). From the acquired images, we measure the instantaneous position and rolling posture of the sphere, the maximum travel distance  $L$, and penetration depth $\delta$.

\section{RESULTS} \label{sec:results}
\begin{figure}
\includegraphics[width=.9\columnwidth]{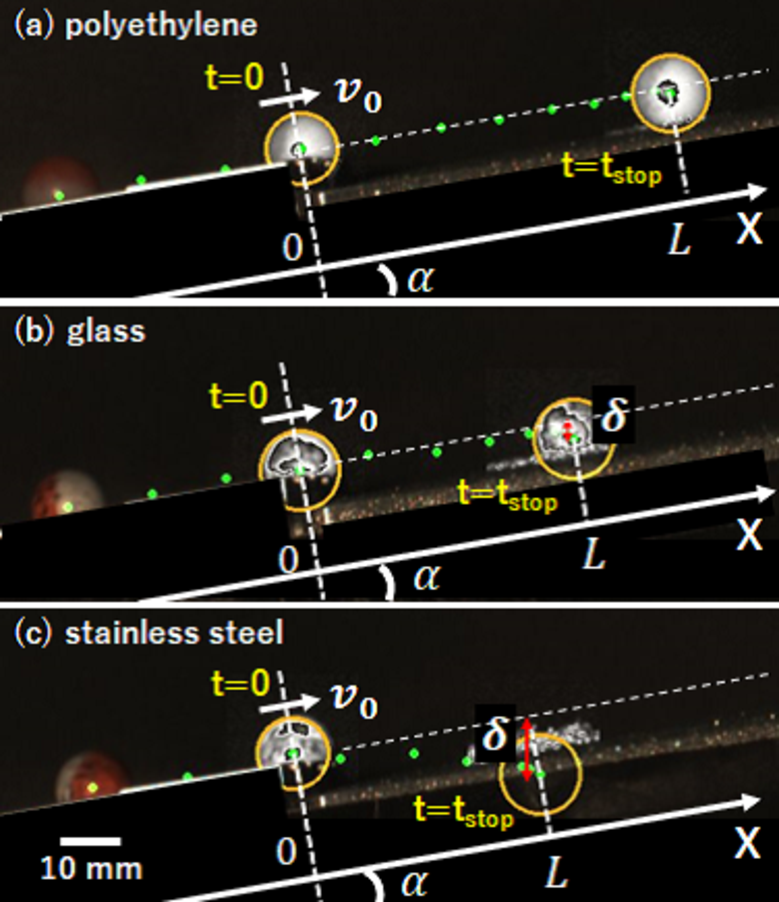}
  \caption{The acquired images of the sphere's motions. (a)~Polyethylene sphere case: $\alpha=10.0^{\circ}$, $v_0=0.52$~m/s, $L=6.0$~cm, and $\delta=1.01$~mm. (b)~Glass sphere case: $\alpha=9.72^{\circ}$, $v_0 = 0.43$~m/s, $L = 4.6$~cm, and $\delta = 3.68$~mm. (c)~Stainless steel sphere case: $\alpha=9.60^{\circ}$, $v_0 = 0.53$~m/s, $L = 4.0$~cm, and $\delta = 11.7$~mm. Yellow circles indicate the circular components identified by the image analysis at $t=0$ and $t=t_\mathrm{stop}$. The green dots indicate the centroid position of the sphere motion every $25$~ms.
}
\label{fig:1005position}
\end{figure}

\begin{figure}
    \centering
    \includegraphics[width=.9\columnwidth]{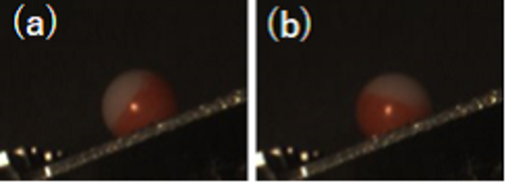}
    \caption{The snapshot images of polyethylene sphere rolling up a slope 
 of $\alpha\simeq20^{\circ}$. The hemispherical part of the sphere is colored in red to measure the rolling posture. The temporal difference between (a) and (b) is 5 ms.}
    \label{pic:pic_roll}
\end{figure}

Fig.~\ref{fig:1005position} shows the 
images of the sphere rolling up a granular slope (actual movies can be found in supplementary movies~\cite{SM}). By detecting the center of the moving sphere, kinematic data (instantaneous centroid position of the sphere) can be measured. The 
precise value of the slope angle $\alpha$ is also measured by the sphere's motion. Specifically, $\alpha$ is measured by the slope of the fitting line of positions of the sphere's motion (in the range of $X < 0$). 
The maximum travel distance  $L$ is defined as the distance between $X=0$ and the position at which the sphere's translational velocity becomes zero. The penetration depth $\delta$ is defined by the vertical sinking distance at the final state.  
We also measure the dynamics of the rolling motion of the sphere. The half of the sphere is colored as shown in Fig.~\ref{pic:pic_roll}. The rolling posture $\theta$ ($\theta=0$ at $t=0$) of the colored hemisphere is measured with a resolution of 0.017 rad. This measurement resolution is smaller than other error factors. By simply differentiate $\theta(t)$ data, we measure the angular velocity $\omega=d\theta/dt$. 

When the stainless steel sphere enters the granular layer, it shows significant penetration. The resultant prominent splashing prevents us from the precise identification of the sphere's position as shown Fig.~\ref{fig:1005position}(c). Therefore, we cannot analyze the time-resolved dynamics for the stainless steel sphere. In the dynamic analysis, we use the data except for the stainless steel sphere.

\subsection{Penetration depth}\label{sec:depth}
\begin{figure}  
    \centering
    \includegraphics[width=\columnwidth]{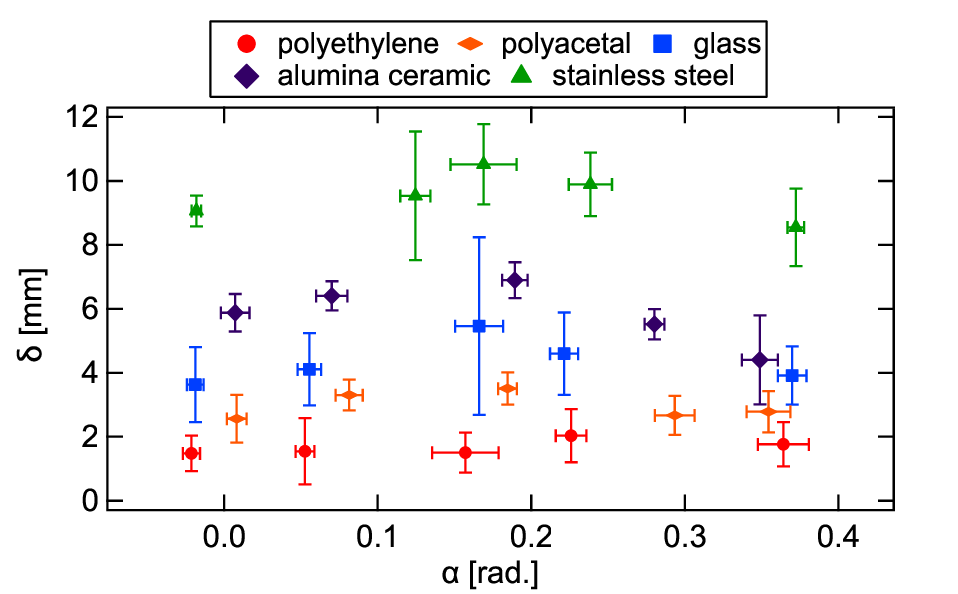}
    \caption{The relation between penetration depth $\delta$ and inclination angle $\alpha$.  Error bars indicate the standard deviation of various initial velocity
    cases and five times repeated experimental runs. One can confirm that $\delta$ does not show clear dependence on $\alpha$. The $v_0$ dependence is also limited in the range of error bars. The penetration depth mainly depends on the density ratio $\rho_s/\rho_g$. 
    }
    \label{fig:depva}
\end{figure}

\begin{figure} 
    \centering
    \includegraphics[width=\columnwidth]{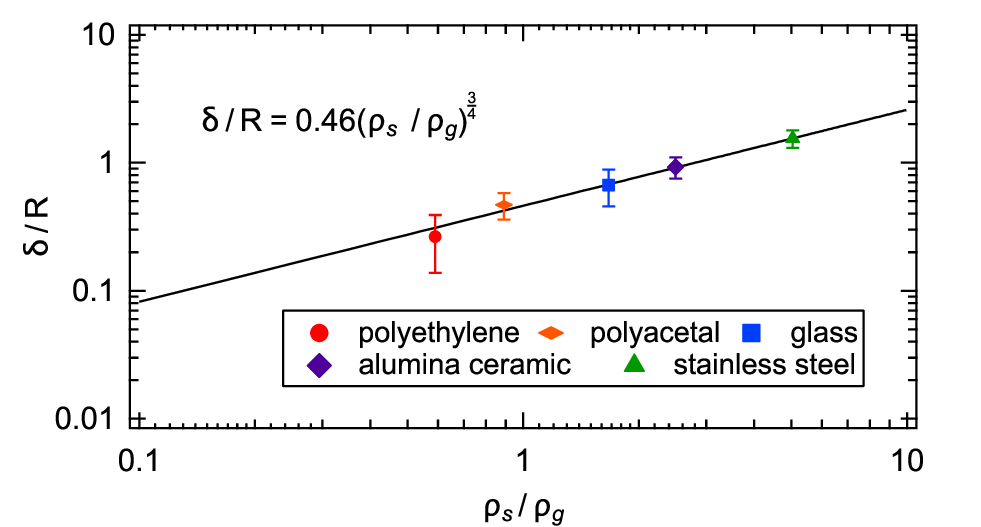}
  \caption{The double logarithm plot of $\delta/R$ vs $\rho_s/\rho_g$. The solid line indicates the scaling relation [Eq.~\eqref{eq:d_scaling}] with $C_{\rho}=0.46$. The value of $C_{\rho}$ is computed by the least-square fitting. Error bars indicate the standard deviation of all the same $\rho_s/\rho_g$ data (with various $v_0$ and $\alpha$ cases).
}
    \label{fig:fitdepth}
\end{figure}

First, we characterize the penetration depth of the sphere. By the image analysis of the sphere's motion, final penetration depth $\delta$ can be extracted even for the stainless steel sphere case (as long as the sphere can be detected in the image). The measured $\delta$ values for various spheres are plotted in Fig.~\ref{fig:depva}. As seen in Fig.~\ref{fig:depva}, $\delta$ value is almost independent of $\alpha$ (and $v_0$) at least in the range of our experimental conditions. The variation of $\delta$ seems to originate from the density difference. Actually, density-dependent penetration depth of a sphere into a granular layer has been investigated in previous works~\cite{Uehara,Melo}. In these studies, the penetration depth of a sphere into a horizontal or inclined granular surface with zero impact velocity was systematically measured. They experimentally found that $\delta$ was scaled by the density ratio between sphere and granular layer, $\rho_s/\rho_g$. Specifically, they proposed a scaling relation,
\begin{equation}
  \frac{\delta}{R}= C_{\rho} \left( \frac{\rho_s}{\rho_g} \right)^{3/4},
  \label{eq:d_scaling}
\end{equation}
where $C_{\rho}$ is a dimensionless constant. The value of $C_{\rho}$ was estimated as $0.51$ for the penetration into a horizontal granular surface~\cite{Uehara} while $C_{\rho}=0.61$ was obtained when the sphere was rolling down the granular slope~\cite{Melo}. To check the applicability of this scaling [Eq.~\eqref{eq:d_scaling}] to the case of sphere rolling up a granular slope, the relation between $\delta/R$ and $\rho_s/\rho_g$ obtained in this study is plotted in Fig.~\ref{fig:fitdepth}. As expected, the data obtained in this experiment follows the scaling of Eq.~\eqref{eq:d_scaling} with $C_{\rho}=0.46$. 

The data agrees well with the scaling relation. Therefore, we consider that the vertical penetration depth is independent of the surface inclination angle $\alpha$ and the motion in $X$ direction. Although this $\delta$ behavior is natural, how this $\delta$ scaling affects the entire rolling-up motion is not a trivial problem. Thus, we carefully analyze translational and rotational motions as well.

\subsection{Translational motion}\label{sec:translational_motion}
\begin{figure*} 
\includegraphics[width=.9\textwidth]{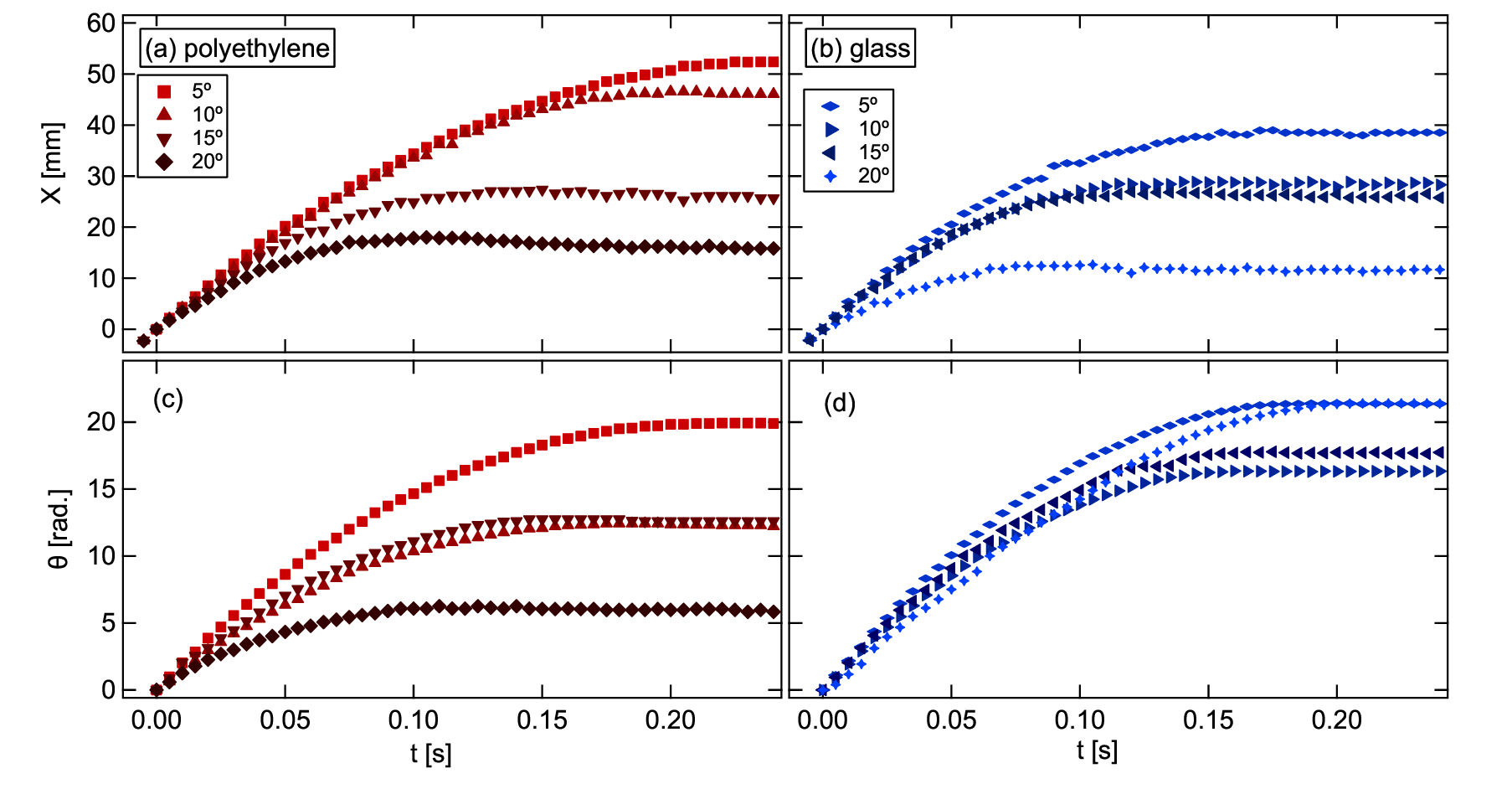}
\caption{(a) and (b) : The translational $X$ position, (c) and (d) : the rotational posture $\theta$ of the sphere rolling up granular slopes are shown as functions of time. The panels (a) and (c) correspond to polyethylene sphere data and the panels (b) and (d) correspond to glass sphere data. The initial velocity for all data shown in this plot is $v_{0} \simeq 0.45$~m/s.}
\label{fig:timepass}
\end{figure*}

Next, we analyze the translational motion of the sphere rolling up a granular slope. In Fig.~\ref{fig:timepass}, the instantaneous position in $X$ direction, $X(t)$, for (a)~polyethylene sphere and (b) glass sphere are displayed. Note that $X(t)$ of the stainless steel spheres cannot be measured in many cases due to the deep penetration. Therefore, we analyze the data of polyethylene, polyacetal, glass, and alumina ceramic spheres in the following analysis. And, the data of polyethylene and glass are mainly plotted as representative examples. Although the data of polyacetal and alumina ceramic spheres are not shown in Fig.~\ref{fig:timepass}, they also show the similar tendency. As clearly seen in Fig.~\ref{fig:timepass}(a) and (b), $X(t)$ seems to approach the asymptotic value (maximum travel distance) $L$. Obviously, $L$ is a decreasing function of inclination angle $\alpha$. 

\begin{figure*} 
    \includegraphics[width=.9\textwidth]{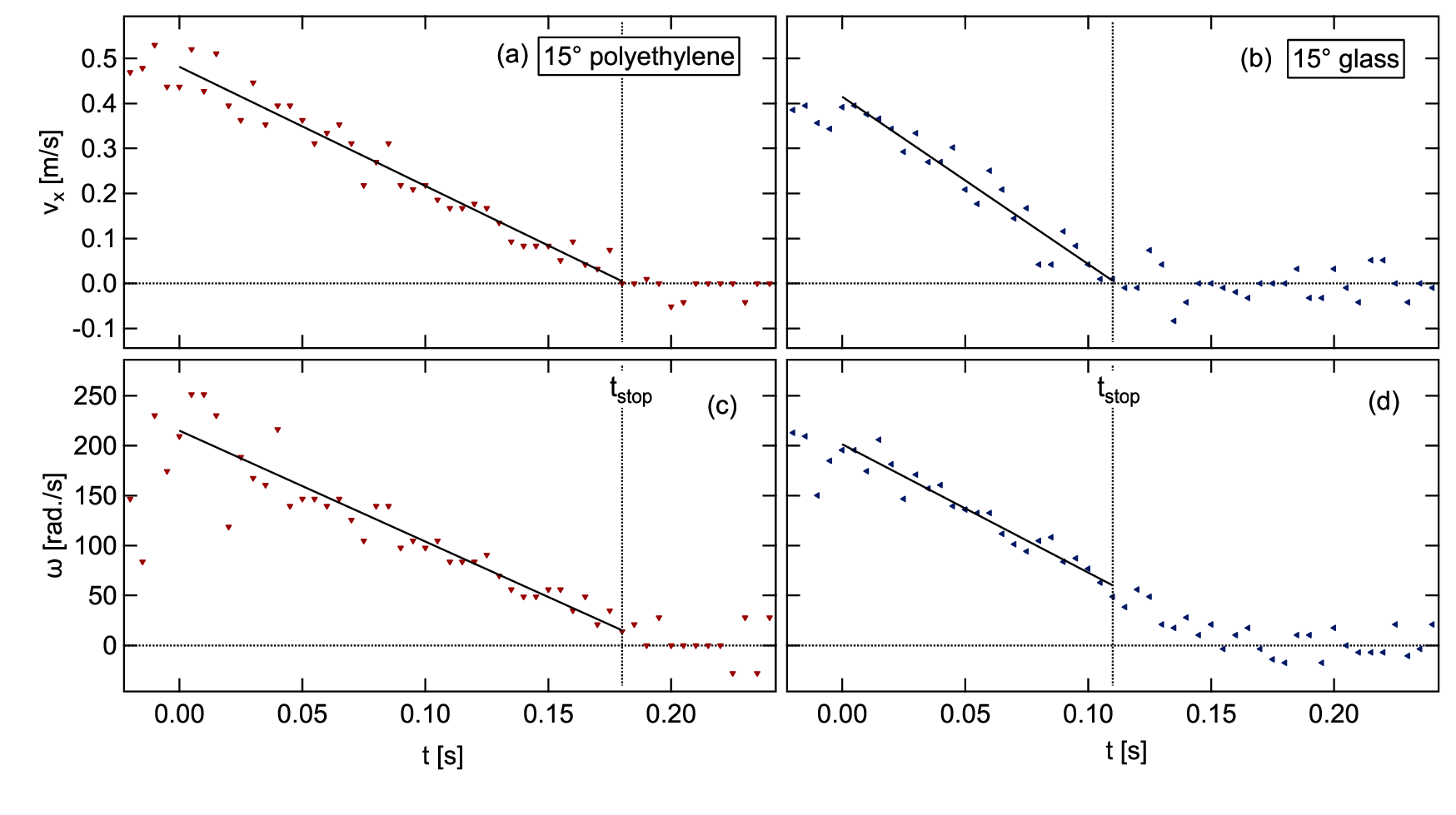}
\caption{(a) and (b) : The translational velocity $v_X$ as a function of time $t$ at $\alpha\simeq 15^{\circ}$[(a) polyethylene sphere and (b) glass sphere]. By the least-square fitting to the linear deceleration (from $t=0$ to the stopping time $t_\mathrm{stop}$ at which $v_{X}(t_\mathrm{stop})=0$), almost constant deceleration $a_X$ can be obtained. (c) and (d) : The angular velocity $\omega(t)$ as a function of time $t$ at $\alpha\simeq15^{\circ}$[(c) polyethylene sphere and (d) glass sphere]. By the least-square fitting to the linear deceleration (from $t=0$ to the stopping time $t_\mathrm{stop}$), almost constant deceleration $\dot{\omega}$ can be obtained.}
\label{fig:t_resolution}
\end{figure*}

\begin{figure} 
\includegraphics[width=\columnwidth]{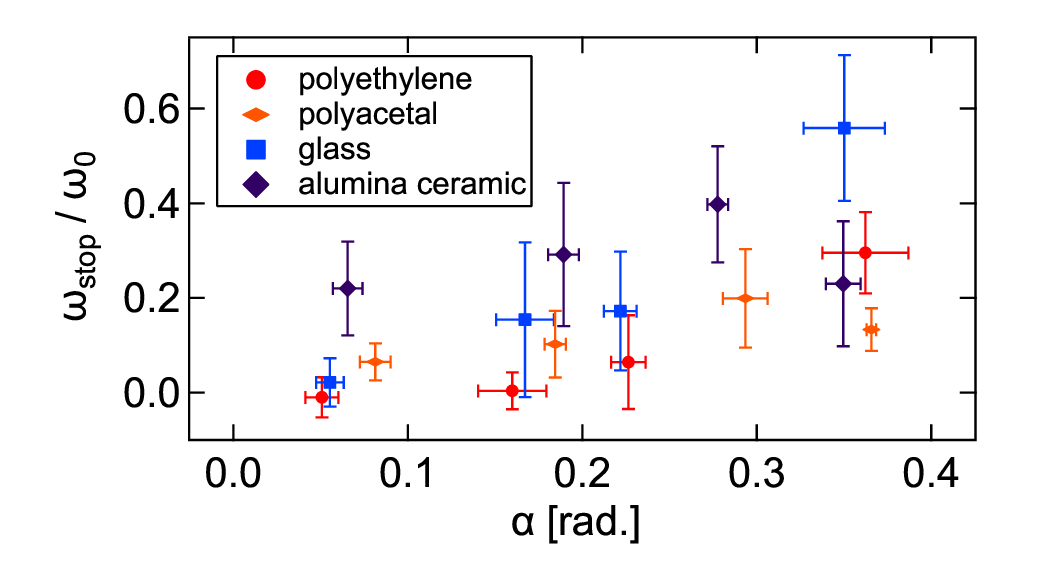}
\centering
\caption{
Measured $\omega_\mathrm{stop}/\omega_{0}$ is plotted as a function of slope angle $\alpha$. Error bars indicate the standard deviation of the data with various $v_0$ cases. $\omega_\mathrm{stop}/\omega_{0}<0$ means the sphere's rotation originating from the backward translational motion after rotational cessation.
}
\label{fig:wstop}
\end{figure}

To characterize the deceleration dynamics, instantaneous velocity $v_X(t)$ is computed by differentiating $X(t)$ data. Typical examples of the $v_X(t)$ data are shown in Fig.~\ref{fig:t_resolution}(a) and (b). As observed in Fig.~\ref{fig:t_resolution}(a) and (b), constant deceleration can be confirmed both in polyethylene sphere and glass sphere. 
From the least square fitting of these data to the linear function, we estimate the average acceleration $a_X$. 
The linear trend in $v_X(t)$ behavior can be observed also in all other experimental data. Therefore, we simply assume $a_X$ is constant during the rolling-up process. Actually, more or less similar constant-acceleration behavior was confirmed also in the rolling down experiment~\cite{Melo}. As far as the penetration depth is shallow, the constant acceleration/deceleration seems to be a reasonable approximation to analyze the sphere's motion on the granular slope. However, physical understanding of the constant $a_X$ has not been provided so far and the origin of this constant $a_X$ is still unknown. Although the microscopic (grain-scale) origin of the constant $a_X$ cannot be easily answered, it can be phenomenologically understood on a macroscopic scale. The constant $a_X$ implies that the force exerting on the sphere is constant. To consider the specific energy balance that determines the constant $a_X$ value, we discuss on friction in Sec.~\ref{sec:analysis}.

\subsection{rolling motion}
Finally, we analyze the rolling motion.
In Fig.~\ref{fig:timepass}(c) and (d), instantaneous rolling posture $\theta(t)$ is displayed. Here, the posture $\theta(t)$ indicates the inclination of boundary (shown in Fig.~\ref{pic:pic_roll}) driven by rotation in clockwise direction.
To characterize the deceleration dynamics of rotation, instantaneous $\omega(t)$ is computed by differentiating $\theta(t)$ data. Typical examples of $\omega(t)$ are shown in Fig.~\ref{fig:t_resolution}(c) and (d). From the least square fitting of these data to the linear function (from $t=0$ to $t_\mathrm{stop}$), we estimate the average
angular acceleration $\dot\omega$, where $t_\mathrm{stop}$ is time at which $v_X = 0$. Similar to the translational motion, constant $\dot\omega$ tendency can be confirmed in all experimental data.

By careful inspection of the measured rolling-up movies, we realize that the translational and rolling motions do not halt simultaneously (see the movies in supplemental materials~\cite{SM}). This behavior can be clearly confirmed also in Fig.~\ref{fig:t_resolution}(c) and (d). In some cases, rolling motion lasts longer than translational motion. We define this behavior as a stuck phenomenon. Indeed, $\omega_\mathrm{stop} = \omega(t_\mathrm{stop})$ is not zero in Fig.~\ref{fig:t_resolution}(d). In order to clearly show this trend, $\omega_\mathrm{stop}/\omega_{0}$ is
measured and plotted in Fig.~\ref{fig:wstop}.
One can confirm that $\omega_\mathrm{stop}/\omega_{0}$ shows an increasing trend with $\alpha$.
We consider $\omega_\mathrm{stop}/\omega_{0}$ is
an indicator characterizing the degree of stuck phenomenon. 

According to Ref.~\cite{deformation}, when light cylinders rolled on a flat granular bed, the cylinders tended to roll backwards before they completely stopped. In general, translational and rolling motions on granular surfaces do not halt simultaneously. In this experiment, we also observe a time lag between translational and rotational cessations. Both the positive and negative $\omega_\mathrm{stop}/\omega_0$ can be confirmed. The former indicates stuck phenomenon and the latter corresponds to the rolling back motion. To understand these peculiar behaviors, energy dissipation due to friction is considered in the next section.

\section{ANALYSIS}\label{sec:analysis}
To quantitatively characterize the observed behaviors, we consider two types of energy dissipation models: (i) energy dissipation by granular deformation during translational motion and (ii) energy dissipation due to the slipping friction during rolling motion. 

\subsection{Energy dissipation due to deformation of the granular layer } \label{sec:energy_dissipation}
To evaluate energy dissipation due to granular deformation by translational motion of the sphere, energy balance should be considered. Here, we consider a simple energy conservation law between two states: $X=0$ and $L$. The energy conservation can be written as,
\begin{equation}
\frac{1}{2}Mv_{0}^2 = Mg(L\sin\alpha - \delta) + F_{d}L, 
  \label{eq:transenergy_conservation}
\end{equation}
where the first term of the right-hand side represents the potential energy. To balance the energy budget, the general dissipative term is introduced in the second term of the right-hand side. For the energy dissipation, we simply assume a constant dissipative force $F_d$. Moreover, $F_d$ can be expressed by the form, 
\begin{equation}
F_{d} = \mu_\mathrm{d}Mg\cos\alpha,
  \label{eq:transF_d} 
\end{equation}
with an effective friction coefficient $\mu_\mathrm{d}$. For the sake of simplicity, we assume $\mu_\mathrm{d}$ is constant in this study. Note that this $\mu_\mathrm{d}$ includes the energy dissipation effects such as deformation of the granular surface.

From Eqs.\eqref{eq:transenergy_conservation} and \eqref{eq:transF_d}, a simple form to estimate the maximum travel distance  $L$ can be obtained as,
\begin{equation}
L= \frac{v_{0}^2 + 2g\delta}{2g(\sin\alpha+\mu_\mathrm{d}\cos\alpha)}. 
  \label{eq:L_form}
\end{equation}

\begin{figure*}
\includegraphics[width=.9\textwidth]{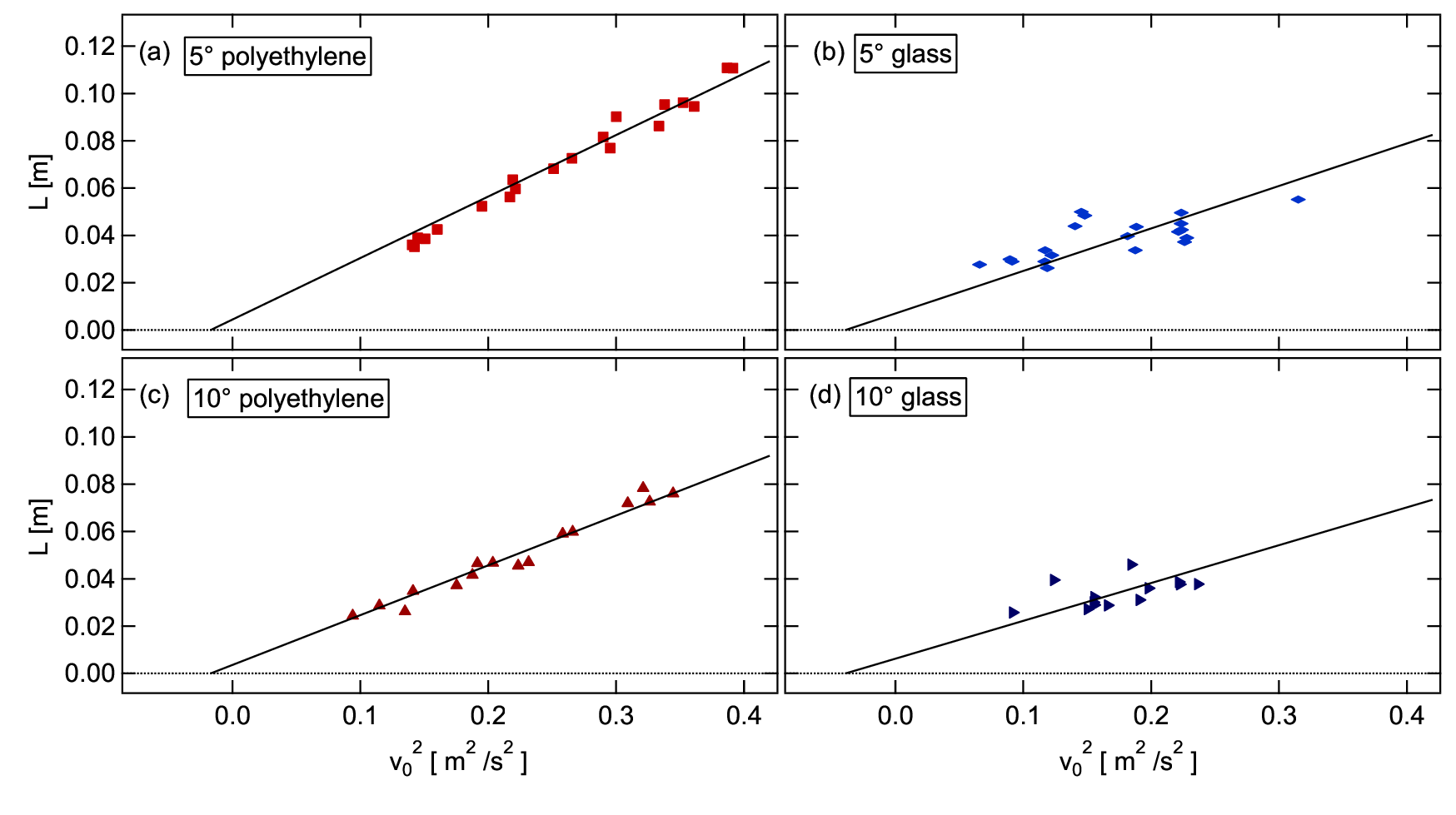}
  \caption{The relations between $L$ and $v_0^2$ are plotted. The solid lines denote the fitting through the fixed point $(v_0^2, L)=(-2g\delta, 0)$. The value of $\delta$ is computed based on Eq.~\eqref{eq:d_scaling}. The positive $L$ value at $v_0=0$ comes from the effect of inclination and the definition. (a)~$5^{\circ}$ polyethylene sphere, (b)~$5^{\circ}$ glass sphere, (c)~$10^{\circ}$ polyethylene sphere, and (d)~$10^{\circ}$ glass sphere. From the gradient of the fitting line, $\mu_\mathrm{d}$ can be estimated using Eq.~\eqref{eq:L_form}.
}
\label{fig:Lfit}
\end{figure*}

Fig.~\ref{fig:Lfit} shows the experimental result of the relation between $L$ and $v_0^2$ for polyethylene and glass spheres at $\alpha\simeq 5^{\circ}$ and $10^{\circ}$. As can be seen, data are consistent with Eq.~\eqref{eq:L_form}. Solid lines in Fig.~\ref{fig:Lfit} indicate the fitting to Eq.~\eqref{eq:L_form}. In the fitting, Eq.~\eqref{eq:d_scaling} is substituted into Eq.~\eqref{eq:L_form}. Then, the intercept of the fitting line is fixed [$(v_{0}^2, L)=(-2g\delta, 0)$]. The linear relation between $L$ and $v_0^2$ supports that the sole fitting parameter in this energy conservation equation, $\mu_\mathrm{d}$, can be regarded as a constant that is independent of $v_0$ (and $\omega_0$). This linear trend is confirmed also in all other experimental results. From the fitting of all $L(v_0^2)$ data (with various $\alpha$ and $\rho_\mathrm{s}$ cases), we compute $\mu_\mathrm{d}$ values. The obtained relation between $\mu_\mathrm{d}$ and $\alpha$ is shown in Fig.~\ref{fig:mua}(a). $\mu_\mathrm{d}$ is an almost constant value on $\alpha$.

\subsection{Energy dissipation due to the rolling motion} \label{sec:rolling}
\begin{figure} 
\includegraphics[width=.95\columnwidth]{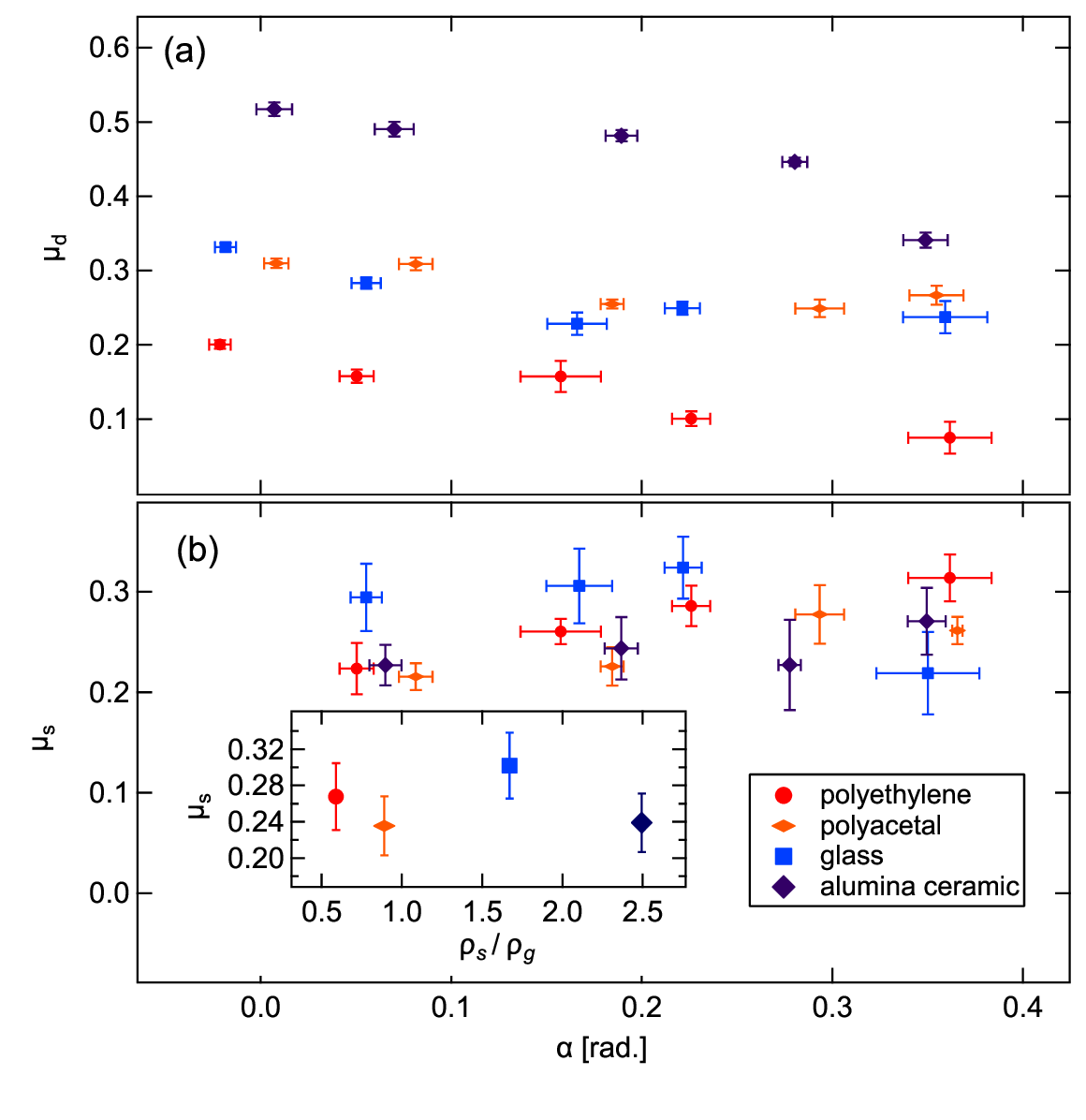}
\centering
\caption{
(a) Friction coefficient characterizing translational energy dissipation $\mu_\mathrm{d}$ and (b) friction coefficient due to the slipping $\mu_\mathrm{s}$ are plotted vs. $\alpha$.  Error bars indicate the standard deviation of the data with various $v_0$ cases and repeated experiments. In the inset of (b), $\mu_\mathrm{s}$ are plotted vs $\rho_s/\rho_g$. Error bars indicate the standard deviation of the all the same $\rho_s/\rho_g$ data (with various $v_0$ and $\alpha$ cases).
}

\label{fig:mua}
\end{figure}

To evaluate dissipation by slipping,
a simple rotational equation of motion is considered,
\begin{equation}
    I\dot{\omega} = -R\mu_\mathrm{s}Mg\cos\alpha,
    \label{eq:roll_eq_motion}
\end{equation}
where $I = (2/5)MR^2$ is the moment of inertia of the homogeneous sphere of mass $M$ and radius $R$, and $\mu_\mathrm{s}$ is the slipping friction coefficient. 
Namely, we assume that the deceleration of the rolling motion can be modeled by the simple slipping torque characterized by the constant friction coefficient $\mu_\mathrm{s}$. Then, the value of $\mu_\mathrm{s}$ can be computed by Eq.~\eqref{eq:roll_eq_motion} and measured $\dot\omega$ (other parameters are known).
The obtained relation between $\mu_\mathrm{s}$ and $\alpha$ is shown in Fig.~\ref{fig:mua}(b) and the relation between $\mu_\mathrm{s}$ and $\rho_\mathrm{s}/\rho_\mathrm{g}$ is shown in the inset of Fig.~\ref{fig:mua}(b). It seems that $\mu_\mathrm{s}$ is independent of $\alpha$ and $\rho_\mathrm{s}$/$\rho_\mathrm{g}$. From the obtained data, typical value of $\mu_\mathrm{s}$ is computed as $\mu_\mathrm{s} = 0.26 \pm 0.04 $.

\section{DISCUSSION}\label{sec:discussion}
\begin{figure} 
    \includegraphics[width=.98\columnwidth]{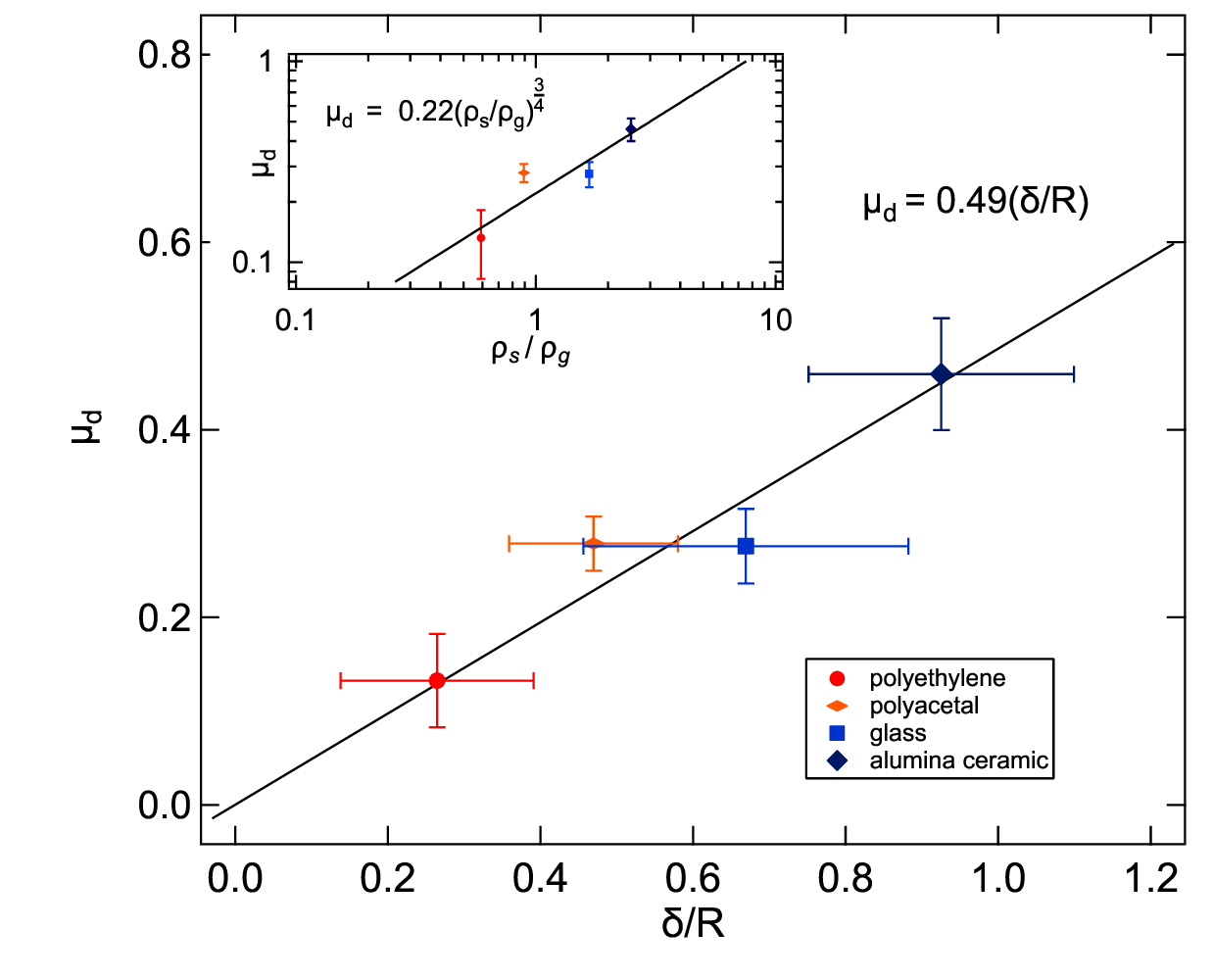}
    \caption{The relation between $\mu_\mathrm{d}$ and $\delta/R$ is plotted. By the least-square fitting, the proportional relation between $\mu_\mathrm{d}$ and $\delta/R$ is obtained. Error bars indicate the standard deviation of all the same $\rho_s/\rho_g$ data (with various $v_0$ and $\alpha$ cases). In the inset, the double logarithm plot of $\mu_\mathrm{d}$ vs $\rho_s/\rho_g$ is shown. The solid line indicates scaling relation.}
    \label{fig:mud_rho}
\end{figure}

In Fig.~\ref{fig:mua}(a), we can confirm $\mu_\mathrm{d}$ is an increasing function of $\rho_s/\rho_g$. This tendency is similar to the relation between $\delta/R$ and $\rho_s/\rho_g$ shown in Fig.~\ref{fig:fitdepth}. As a result, we find a proportional relation, 
\begin{equation}
    \mu_\mathrm{d} = C_\mathrm{d}\frac{\delta}{R},  
    \label{eq:mud_delta}
\end{equation}
where $C_\mathrm{d} = 0.49$ is a fitting parameter (Fig.~\ref{fig:mud_rho}). Since $\delta/R$ is scaled by $(\rho_s/\rho_g)^{3/4}$ [Eq.\ref{eq:d_scaling}], this relation is equivalent to $\mu_\mathrm{d} \sim (\rho_s/\rho_g)^{3/4}$ shown in the inset of Fig.~\ref{fig:mud_rho}. In general, drag force can be proportional to the contacting or cross-sectional area of the moving object. For example, Pacheco-Vázquez and Ruiz-Suárez revealed the linear relation between the drag force of an intruder into a granular medium and a cross-sectional area~\cite{F.Pacheco:2009}. In this regard, Eq.~(\ref{eq:mud_delta}) probably means the dissipative granular drag force is proportional to the contact area ($\sim \delta R$) in the current experimental setup. In the translational motion, grains of contacting cross-sectional region have to be removed. Therefore, we consider $\mu_\mathrm{d}\sim \delta$. However, $\mu_\mathrm{s}$ is independent of $\delta$, 
and the actual slipping surface could be independent of $\delta$. To further discuss the issue, precise measurements of the contacting area and deformed groove are necessary.

When the sphere's density is very large, deep penetration will be observed and the precise measurement of the sphere's motion becomes quite difficult. In such a situation, even the qualitative behavior might show intrinsically different tendency. In this sense, the frictional characterization discussed so far can be applicable only to the shallow penetration case.

Furthermore, the measured friction coefficients cannot be directly compared with the bulk friction coefficients (as material properties) that are based on the Amontons-Coulomb law~(e.g.,~\cite{Kawamura:2012}). Of course, we consider that the granular friction defined and measured in this study must be different from such conventional ones although all the friction coefficients can be regarded as certain constants under the current experimental conditions. Much more systematic comparison should be performed in the future study. 

In addition, there are some restrictions in the current experiment. For example, we do not vary the surface frictional property of the sphere. The experiment with rough-(frictional)-surface spheres is an interesting future problem. Variations of surface property could affect the value of $\mu_\mathrm{d}$ and $\mu_\mathrm{s}$. Besides, since the sphere is accelerated by rolling over the rail, $v_0$ and $\omega_0$ are not independent in this experiment. To fully characterize the general stuck phenomena, experiments with independently controlled $v_0$ and $\omega_0$ should be performed. The shape parameter is also a possible key parameter. Espinosa et al. revealed that an imperfect body induces nontrivial motion~\cite{Macros:2023}. The actual surface of the vehicles wheel is not perfectly smooth unlike our research. This point is also a crucial future issue.

Extension of this type of research to the case of the externally driven object (like vehicle wheeling case) would also be important for the practical application to the terramechanics issues. Terramechanics model mainly considers steady motion (without deceleration) although the stuck frequently occurres in actual driving scenes. Therefore, we believe our experimental findings based on changing $v(t)$, $\omega(t)$ are crucial.

Finally, we briefly discuss the application of the current result to the prevention of stuck phenomena related to friction. Since the current experimental system is not driven by torque and normal loading, it is difficult to directly apply our findings to actual stuck problem in vehicle driving. However, according to Fig.~\ref{fig:wstop}, the degree of slipping rotation after the translational cessation increases as the $\alpha$ increases. This means that the risk of stuck occurrence (without any driving accelerator) increases when the sphere tries to roll up the steeper slope. To reduce the degree of stuck, $\alpha$ should be small. In this study, the degree of stuck motion for the passively rolling sphere is quantitatively measured for the first time.

Besides, by using the empirical law, $\mu_\mathrm{d} = 0.49(\delta/R)$, we can predict sphere motion at any time. In specific, we can predict the maximum travel distance  $L$ by using Eqs.~\eqref{eq:d_scaling} and ~\eqref{eq:L_form} in arbitrary initial conditions. As discussed so far, the sphere is rolling up with a constant deceleration $a_X$, until $t=t_\mathrm{stop}$, satisfying $v(t) = v_0 - \frac{v_0^2}{2L}t$. Regarding the rotational motion, $\omega(t)$ can be computed by Eq.~(\ref{eq:roll_eq_motion}) with $\mu_\mathrm{s}(\simeq 0.26 \pm 0.04)$. Then, the sphere's motion both in translational and rotational directions can be completely computed. That is, in this study, we find some useful relations for the sphere's motion on a granular slope by obtaining parameters $\mu_\mathrm{d}$ and $\mu_\mathrm{s}$.

\section{CONCLUSION} \label{sec:conclusion}
This study reveals the dynamics of a sphere rolling up a granular slope. By systematically varying the initial velocity $v_0$ (and initial angular velocity $\omega_0$), angle of slope $\alpha$, and density ratio between the sphere and granular layer $\rho_s/\rho_g$, the sphere's motion was measured and analyzed. First, the penetration depth of the sphere during the rolling-up process was measured and scaled as $\delta/R =C_{\rho}(\rho_s/\rho_g)^{3/4}$. This scaling is consistent with previous studies of the sphere penetration into a granular layer. Next, translational and rotational motions of the rolling-up sphere are measured and characterized as constant deceleration dynamics. Then,
 we reveal that $\omega_\mathrm{stop}$ increases with $\alpha$.
To characterize the sphere's motion, we consider two kinds of friction coefficients, $\mu_\mathrm{d}$ and $\mu_\mathrm{s}$. The former and latter represent friction due to the deformation of granular layer and slipping friction, respectively. 
We obtain the relation, $\mu_\mathrm{d} = 0.49(\delta/R)$, meaning the contact area between the sphere and granular layer is a key factor of translational energy dissipation. The $\mu_\mathrm{s}$ is almost independent of $\alpha$ and $\rho_\mathrm{s}/\rho_\mathrm{g}$, showing a constant value $\mu_\mathrm{s}\simeq 0.26 \pm 0.04$. From the system parameters, $\rho_\mathrm{s}/\rho_\mathrm{g}$, $R$ and $\alpha$, we can estimate $\delta$ and $\mu_\mathrm{d}$. Then, by considering constant $\mu_\mathrm{s}$ and initial conditions $v_0$ ($\omega_0$), we can completely predict the sphere's rolling-up motion. Since the range of $\rho_s/\rho_g$ variation is limited in this study, more systematic experiments with wider parameter ranges are crucial future problem. 

\section*{Acknowledgements}
We thank JSPS KAKENHI for financial support under Grant No.~18H03679. This work was partially supported also by the research grant of The Information Center of Particle Technology, Japan (2023), and JSPS-DST Bilateral Program, Grant No.~JPJSBP120227710.

\bibliography{fuk}

\end{document}